\documentclass{ifacconf}

	\usepackage{mathtools}
	\usepackage{algorithm}
	\usepackage{url}
	\usepackage{xcolor}
	\usepackage{color}
	\usepackage{amsmath}
	\usepackage{caption}
	\usepackage{subcaption}
	\usepackage{setspace}
	\usepackage{graphicx}	
	\usepackage{setspace}
	\usepackage{textcomp}
	\usepackage{setspace}
	\usepackage{verbatim} 
	\usepackage{amssymb} 
	\graphicspath{{figures/}}

	\usepackage{cite}
	
	\mathchardef\mhyphen="2D

\usepackage{graphicx}      
\usepackage{natbib}        
\begin{document}
\begin{frontmatter}

\title{State Prediction of \\ Human-in-the-Loop Multi-rotor System \\ with Stochastic Human Behavior Model\thanksref{footnoteinfo}} 

\thanks[footnoteinfo]{The authors would like to acknowledge that this work is supported by NSF CNS-1836952. \\ This work has been submitted to IFAC for possible publication.}

\author[First]{Joonwon Choi} 
\author[First]{Sooyung Byeon} 
\author[First]{Inseok Hwang}

\address[First]{Purdue University, West Lafayette, IN 47907 USA \\ (e-mail: \{choi774, sbyeon, ihwang\}@purdue.edu).}

\begin{abstract}                
Reachability analysis is a widely used method to analyze the safety of a Human-in-the-Loop Cyber Physical System (HiLCPS). This strategy allows the HiLCPS to respond against an imminent threat in advance by predicting reachable states of the system. However, it could lead to an unnecessarily conservative reachable set if the prediction only relies on the system dynamics without explicitly considering human behavior, and thus the risk might be overestimated. 
To reduce the conservativeness of the reachability analysis, we present a state prediction method which takes into account a stochastic human behavior model represented as a Gaussian Mixture Model (GMM). 
In this paper, we focus on the multi-rotor in a near-collision situation. The stochastic human behavior model is trained using experimental data to represent human operators' evasive maneuver. Then, we can retrieve a human control input probability distribution from the trained stochastic human behavior model using the Gaussian Mixture Regression (GMR). 
The proposed algorithm predicts the probability distribution of the multi-rotor's future state based on the given dynamics and the retrieved human control input probability distribution. Besides, the proposed state prediction method considers the uncertainty of the initial state modeled as a GMM, which yields more robust performance. Human subject experiment results are provided to demonstrate the effectiveness of the proposed algorithm. 
\end{abstract}


\end{frontmatter}

\section{Introduction}

Ensuring safe operation is one of the most important tasks to enhance the reliability of a Human-in-the-Loop Cyber Physical System (HiLCPS). To this end, there have been various approaches to assess the safety of the HiLCPS. 
Reachability analysis, a technique to compute a set of reachable state of the HiLCPS, has been commonly applied to derive the safety envelope of the system. For instance, a safety guaranteeing controller which considers a human counterpart was developed based on the Hamilton-Jacobi backward reachable set in \cite{leung2020infusing}. Note that, the stochastic reachability analysis can improve further flexibility compared to the conventional reachability analysis: the stochastic reachable set includes the probabilistic information of the state, which is beneficial for making less conservative decisions by ignoring the reachable state with negligible probability to be arrived (\cite{vinod2017forward}).

Incorporating human behavioral data is another option to minimize the conservativeness of the reachable set. It prevents the reachable set from including the redundant state which a human operator does not tend to reach in practice. Nevertheless, there have been relatively few approaches that compute reachable states based on human behavioral data. Except some existing papers (\cite{driggs2018robust, govindarajan2017data}), most of the works relied on the dynamics of a system or a given bound of states. Even most of the stochastic reachability analysis approaches only considered the stochasticity coming from noise or disturbance.
To address this issue, we propose a state prediction method that can reduce the conservativeness of the existing reachability analysis by infusing a stochastic human behavior model. 
We select the Gaussian Mixture Model (GMM) as a basis model. The GMM has been proven to be a good representation of human operators' reaction (\cite{angkititrakul2011modeling, wang2018learning}). Moreover, the GMM can approximate non-Gaussian distributions (\cite{pishdad2016approximate}), thereby allowing our proposed algorithm to be extended for more general stochastic human models with non-Gaussian distributions.

The objective of the proposed algorithm is to predict the future state of the multi-rotor in a near-collision situation. First, the stochastic human behavior model is trained as a GMM using the Expectation Maximization (EM) algorithm and the flight trajectory from human subject experiments. Then, the state of the multi-rotor propagates according to the given linear dynamics and the human operator's control input probability distribution retrieved using the Gaussian Mixture Regression (GMR). 
Assuming the initial uncertainty is also given as a GMM, the resulting probability distribution of the future state can be derived as a form of the GMM. A Gaussian mixture reduction technique is used to check the ever growing number of Gaussian components to alleviate the computational complexity. 
The contributions of this paper are as follows: 
(a) We propose the state prediction method that can explicitly account for the stochastic human behavior. As a result, the probability distribution of the predicted state is obtained, which reduces the conservativeness of the conventional reachable set methods. 
(b) The proposed state prediction algorithm is demonstrated using the stochastic human behavior model of the multi-rotor. The stochastic human behavior model is trained using the data obtained from the human subject experiments setting on the near-collision scenario of the multi-rotor. Through the experiments, the performance of the proposed state prediction algorithm is demonstrated. 

The rest of the paper is organized as follows: In Section 2, detailed information about the experiment scenario and the multi-rotor system is given. The method to train the stochastic human behavior model is introduced in Section 3. In Sections 4 and 5, the state prediction algorithm for the multi-rotor and the experiment results are presented, respectively. Lastly, conclusions are given in Section 6.

\section{Near-collision scenario}

\begin{figure} [t]
   	  \centering
	  \includegraphics[scale=0.4]{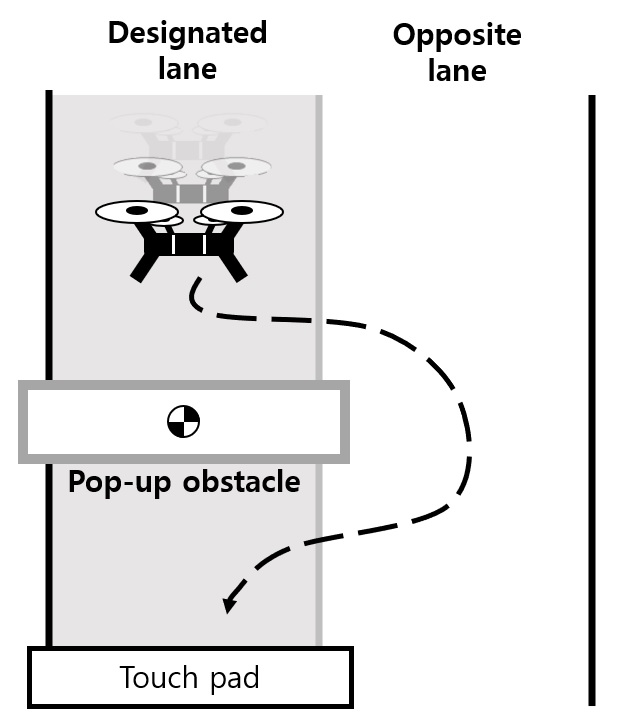}
      \caption{Landing mission of the multi-rotor}
      \label{scenario_fig}
\end{figure}

Figure \ref{scenario_fig} shows a multi-rotor landing mission scenario for human subject experiments. The objective of the experiment is to extract collision avoidance reaction of a human operator in the multi-rotor system. Motivated by the existing works about rear-collision of a car (\cite{luster2021preliminary, angkititrakul2011modeling}), we design the simulator which can observe the participants' evasive maneuver. The simulation is set on the 2-D environment with a pop-up obstacle. 
The participants should land the multi-rotor on a touch pad without any collision with the obstacle. In addition, the participants are asked to follow a designated lane while maintaining constant downward speed during the experiments.
In each trial, the location where the obstacle pops up is not informed to the participants in advance. Thus, the recorded flight data contains the pure reaction of the participants to avoid the obstacle. By focusing on the initial few seconds of the trajectory after the obstacle is spawned, we can train a stochastic human behavior model that represents the participants' collision avoidance reaction.

\begin{figure} [t]
   	  \centering
	  \includegraphics[scale=0.33]{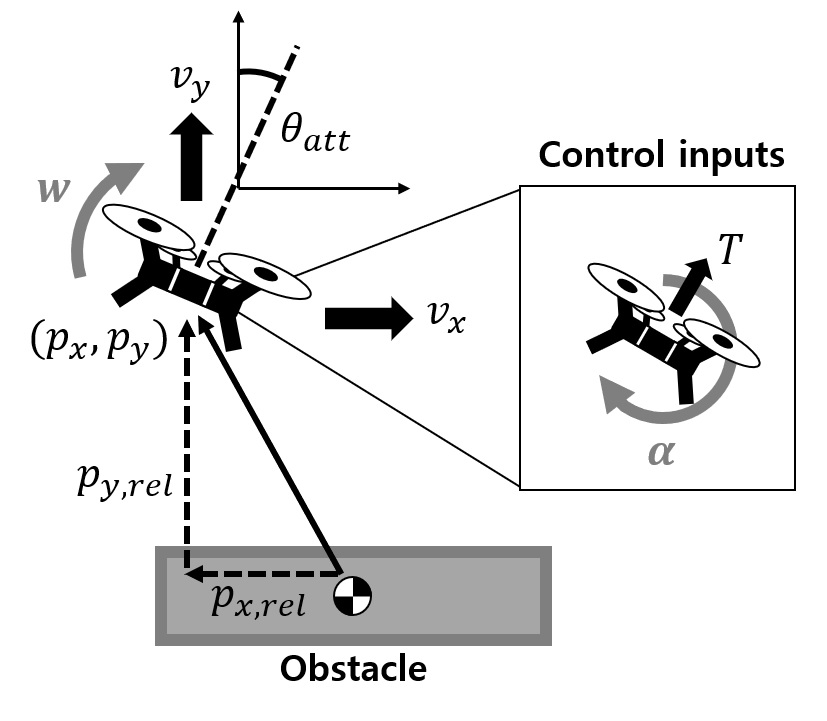}
      \caption{State variables of the multi-rotor}
      \label{sim_var}
\end{figure}

Figure \ref{sim_var} shows the state variables of the simulated multi-rotor. Based on \cite{sabatino2015quadrotor} and \cite{byeon2021skill}, the dynamics of the multi-rotor is modeled as
\begin{equation} \label{dynamics}
\vec{x}_{k+1} = A\vec{x}_k + B\vec{u}_k,
\end{equation}
where $\vec{x}_k$ and $\vec{u}_k$ are the state and the control input vectors at time step $k$, respectively. Here, the state vector $\vec{x}_k \in \mathbb{R}^6$ is defined as $\vec{x}_k = [p_{x,k}, p_{y,k}, \theta_{att,k}, v_{x,k}, v_{y,k}, w_k]^T$, which consists of the position ($p_{x,k}, p_{y,k}$), the attitude ($\theta_{att,k}$), the linear velocity ($v_{x,k}, v_{y,k}$), and the angular velocity ($w_k$). The control input $\vec{u}_k \in \mathbb{R}^2$ is defined as $\vec{u}_k = [\alpha_k, T_k]^T$ where $\alpha$ and $T$ are  the angular acceleration and the thrust, respectively. $A$ and $B$ matrices can be written as 
\begin{equation}
A = I_6 + 
\begin{pmatrix}
0 & 0 & 0 & 1 & 0 & 0 \\
0 & 0 & 0 & 0 & 1 & 0 \\
0 & 0 & 0 & 0 & 0 & 1 \\
0 & 0 & g & 0 & 0 & 0 \\
0 & 0 & 0 & 0 & k_1 & 0 \\
0 & 0 & k_2 & 0 & 0 & k_3 
\end{pmatrix} 
\Delta t,
\end{equation}

\begin{equation}
B =  
\begin{pmatrix}
0 & 0  \\
0 & 0  \\
0 & 0  \\
0 & 0  \\
0 & 1/m  \\
1/I_x & 0  
\end{pmatrix} 
\Delta t,
\end{equation}
where $\Delta t$ is the discretization time interval, $g$ is the gravitational acceleration, $m$ and $I_x$ are the mass and the moment of inertial of the multi-rotor, respectively, $k_1, k_2,$ and $k_3$ are the controller parameters, and $I_6$ is the $6\times6$ identity matrix.

\section{Stochastic human behavior model}

In this section, we train the stochastic human behavior model using the data collected from the human subject experiments. Using the Expectation Maximization (EM) algorithm, we can compute the stochastic human behavior model which contains the joint distribution between the participants' control input and the time elapsed after the obstacle is spawned. The trained stochastic human behavior model will be utilized as the prior knowledge for the multi-rotor's state prediction in Section 4.

The GMM is a combination of multiple Gaussian distributions. Assuming a distribution of state $x$ follows the GMM, the probability distribution of $x$, $P(x)$, is defined as
\begin{equation}
P(x) = \sum^M_{i=1} \pi_{i} N(\mu_{i},\Sigma_{i}),
\end{equation}
where $\pi_i$ is the weight of each Gaussian component satisfying $\sum^M_{i=1} \pi_i = 1$ and $N(\mu_{i},\Sigma_{i})$ is the Gaussian distribution which has $\mu_i$ and $\Sigma_i$ as the mean and the covariance, respectively. In this paper, we focus on the the joint distribution between the human control input and time, which can be represented as a GMM. Accordingly, the trajectory used to train the stochastic human behavior model is written as $\vec{\zeta} = [\vec{\zeta}_0 ^T, \vec{\zeta}_1 ^T, \cdots, \vec{\zeta}_{t_f}^T ]^T$ where $\vec{\zeta}_k = [t_k, \vec{u}_k^T ]^T$, $t_k = k \Delta t$, and $t_f$ is the end time of the trajectory. It is worth noting that $t_0$ is the time when the obstacle is spawned. Thus, the stochastic human behavior model is trained using the trajectory after the obstacle pops up. 
 The proper parameters of the stochastic human behavior model can be computed by feeding $\vec{\zeta}$ to the EM algorithm. The trained stochastic human behavior model can be expressed as (\cite{calinon2016tutorial})
\begin{equation} \label{pilot_gmm}
P(\vec{u},\vec{t}) = \sum^M_{i=1} \pi_{p,i} N(\vec{\mu}_{p,i},\vec{\Sigma}_{p,i}),
\end{equation}
where $\vec{u} = [\vec{u}_1^T, \vec{u}_2^T, \cdots, \vec{u}_{t_f}^T]^T$, $\vec{t} = [t_1, t_2, \cdots, t_{t_f}]^T$, and $\pi_{p,i}$, $\vec{\mu}_{p,i}$, and $\vec{\Sigma}_{p,i}$ are the weight, mean, and covariance matrix for each Gaussian component, respectively. The EM algorithm finds the value of each Gaussian component's parameters ($\pi_{p,i}$, $\vec{\mu}_{p,i}$, and $\vec{\Sigma}_{p,i}$) which are the best fit for representing the given trajectory. 
Throughout the paper, we assume that the stochastic human behavior model is composed of $M$ number of Gaussian components, which is a design parameter. 

\section{State prediction of multi-rotor}

The proposed state prediction algorithm predicts the probability distribution of the $T$-step future state ($\vec{x}_{k+T}$). The initial state propagates through the multi-rotor dynamics with uncertainty and the human input probability distribution retrieved using the Gaussian Mixture Regression (GMR). To reduce the computation burden, a Gaussian mixture reduction method is applied for each iteration. 

\subsection{State prediction based on stochastic human behavior model}

The GMR allows us to retrieve the conditional distribution of the human control input at a given time instant (\cite{calinon2016tutorial, stulp2015many}). From \eqref{pilot_gmm}, let $\vec{\mu}_{p,i}$ and $\vec{\Sigma}_{p,i}$ be defined as
\begin{equation} \label{gmr_eqs}
\vec{\mu}_{p,i} = 
\begin{pmatrix}
\vec{\mu}^u_{p,i}   \\
\vec{\mu}^t_{p,i}  
\end{pmatrix},
\vec{\Sigma}_{p,i} = 
\begin{pmatrix}
\vec{\Sigma}^u_{p,i} & \vec{\Sigma}^{ut}_{p,i}     \\
\vec{\Sigma}^{tu}_{p,i} & \vec{\Sigma}^t_{p,i}  
\end{pmatrix}.
\end{equation}

In \eqref{gmr_eqs}, $\vec{\mu}^u_{p,i} \in \mathbb{R}^2$ is the mean fraction corresponding to the human control input, $\vec{\mu}^t_{p,i} \in \mathbb{R}$ is that of time, and $\vec{\Sigma}^u_{p,i},\ \vec{\Sigma}^t_{p,i},\ \vec{\Sigma}^{ut}_{p,i},$ and $\vec{\Sigma}^{tu}_{p,i}$ are the corresponding covariance fractions, respectively. Then, the conditional distribution of the human input at time step $k$ is (\cite{calinon2016tutorial})
\begin{equation} \label{gmr}
P(\vec{u}| t_k) = P(\vec{u}_k) = \sum^M_{i=1} \hat{\pi}_{p,i}(t_k) N(\hat{\vec{\mu}}_{p,i}(t_k), \hat{\vec{\Sigma}}_{p,i}),
\end{equation}
where
\begin{equation}
\hat{\vec{\mu}}_{p,i}(t_k) = \vec{\mu}^u_{p,i} + \vec{\Sigma}^{ut}_{p,i} \vec{\Sigma}^{t^{-1}}_{p,i}(t_k - \vec{\mu}^t_{p,i}),
\end{equation}
\begin{equation}
\hat{\vec{\Sigma}}_{p,i} = \vec{\Sigma}^u_{p,i} - \vec{\Sigma}^{ut}_{p,i} \vec{\Sigma}^{t^{-1}}_{p,i} \vec{\Sigma}^{tu}_{p,i}, 
\end{equation}
\begin{equation}
\hat{\pi}_{p,i}(t_k) = {\pi_{p,i} N(t_k|\vec{\mu}^t_{p,i},\vec{\Sigma}^t_{p,i}) \over \sum^M_j \pi_{p,j} N(t_k|\vec{\mu}^t_{p,j},\vec{\Sigma}^t_{p,j}) }.
\end{equation}

Our objective is to compute the probability distribution of the future state at time step $k+T$, $P(x_{k+T})$. Assume that the current state ($\vec{x}_k$) has uncertainty represented as a GMM with $L$ Gaussian components:
\begin{equation} \label{current_state}
\vec{x}_k \sim \sum^L_{i=1} \pi_{\vec{x}_k,i} N(\vec{\mu}_{\vec{x}_k,i},\vec{\Sigma}_{\vec{x}_k,i})
\end{equation}

The marginal distribution of $\vec{x}_{k+1}$ is obtained as (\cite{karumanchi2021comparing})
\begin{equation} \label{chapman}
P(\vec{x}_{k+1}) = \int_{\vec{x}_k} P(\vec{x}_{k+1}|\vec{x}_k)P(\vec{x}_k)d\vec{x}_k.
\end{equation}

From \eqref{chapman}, using the dynamics \eqref{dynamics} and the input probability distribution \eqref{gmr}, $P(\vec{x}_{k+1}|\vec{x}_k)$ can be rewritten as:
\begin{equation} \label{(13)}
P(\vec{x}_{k+1}|\vec{x}_k) = \sum^M_{i=1} \hat{\pi}_{p,i}(t_k) N(A\vec{x}_k + B\hat{\vec{\mu}}_{p,i}(t_k), B\hat{\vec{\Sigma}}_{p,i}B^T)
\end{equation}

By substituting equation \eqref{(13)} for $P(\vec{x}_{k+1}|\vec{x}_k)$ in \eqref{chapman}, we obtain
\begin{multline}
P(\vec{x}_{k+1}) = \\ \int_{\vec{x}_k} \sum^M_{i=1} \hat{\pi}_{p,i}(t_k) N(A\vec{x}_k + B\hat{\vec{\mu}}_{p,i}(t_k), B\hat{\vec{\Sigma}}_{p,i}B^T) P(\vec{x}_k)d\vec{x}_k,
\end{multline}
and if we replace $P(\vec{x}_k)$ with \eqref{current_state}, we have
\begin{multline} \label{bf_final}
P(\vec{x}_{k+1}) =  \\  \int_{\vec{x}_k} \sum^M_{i=1}  \hat{\pi}_{p,i}(t_k) N(A\vec{x}_k + B\hat{\vec{\mu}}_{p,i}(t_k), B\hat{\vec{\Sigma}}_{p,i}B^T) \cdot \\ \sum^L_{j=1} \pi_{\vec{x}_k,j} N(\vec{\mu}_{\vec{x}_k,j},\vec{\Sigma}_{\vec{x}_k,j}) d\vec{x}_k.
\end{multline}

By \cite{karumanchi2021comparing}, equation \eqref{bf_final} can be rewritten as a GMM with $ML$ number of Gaussian components:
\begin{multline}  \label{final}
P(\vec{x}_{k+1}) = {1 \over \eta} \sum^M_{i=1}  \hat{\pi}_{p,i}(t_k)  \sum^L_{j=1} \pi_{\vec{x}_k,j} \cdot \\ N(A\vec{\mu}_{\vec{x}_k,j} + B\hat{\vec{\mu}}_{p,i}(t_k), B\hat{\vec{\Sigma}}_{p,i}B^T + A\vec{\Sigma}_{\vec{x}_k,j}A^T),
\end{multline}
where $\eta$ is a normalization constant to ensure the integral of \eqref{final} is 1.

One can compute $P(\vec{x}_{k+T})$ by repeating \eqref{current_state}-\eqref{final} to time step $T$. However, this process is computationally demanding as the number of Gaussian components grows rapidly. For each time step, it increases by $M$ times since we assume the stochastic human behavior model is composed of $M$ Gaussian components. 

\subsection{Gaussian mixture reduction}

As shown in the previous section, the combination of GMMs leads to the rapid growth of the number of Gaussian components. To address this issue, there have been various studies to efficiently reduce the number of Gaussian components (\cite{zhang2020unified}). One of the methods is using the Kullback-Leibler (K-L) divergence, which represents the difference between two distributions. Unfortunately, the closed form of the K-L divergence between two GMMs is known to be unavailable. To tackle this problem, a novel reduction algorithm was proposed in \cite{runnalls2007kullback} based on the upper bound of the K-L divergence. 

Let $\hat{P}(\vec{x}_{k+1})$ be the reduced GMM from $P(\vec{x}_{k+1})$ by merging the Gaussian components $i$ and $j$ into a single Gaussian. The author selected to merge the pair which induces the lowest upper bound of the K-L divergence between $\hat{P}(\vec{x}_{k+1})$ and $P(\vec{x}_{k+1})$. We adopt this approach to reduce the number of Gaussian components. Let $(\vec{\Sigma}_i, \vec{\Sigma}_j)$ be a pair of the covariance matrices of the selected Gaussian components from $P(\vec{x}_{k+1})$. $(\pi_i, \pi_j)$ and $(\vec{\mu}_i, \vec{\mu}_j)$ are the corresponding weights and means. The merged covariance $\vec{\Sigma}_{ij}$ can be computed as
\begin{multline} \label{merge_cov}
\vec{\Sigma}_{ij} = {\pi_i \over \pi_i+\pi_j} \vec{\Sigma}_i + {\pi_j \over \pi_i+\pi_j} \vec{\Sigma}_j + \\ {\pi_i \pi_j \over (\pi_i+\pi_j)^2}(\vec{\mu}_i - \vec{\mu}_j)(\vec{\mu}_i - \vec{\mu}_j)^T. 
\end{multline}

From \eqref{merge_cov}, the upper bound of the K-L divergence ($KL_{up}(i,j)$) is defined as follow:
\begin{multline}
KL_{up}(i,j)= {1 \over 2}[ (\pi_i + \pi_j)ln(det(\vec{\Sigma}_{ij})) - \\ \pi_i ln(det(\vec{\Sigma}_i))  - \pi_j ln(det(\vec{\Sigma}_j)) ]
\end{multline}

Then, the pair $(i,j)$ with the minimum $KL_{up}$ is merged into a single Gaussian, $\pi_{ij} N(\vec{\mu}_{ij}, \vec{\Sigma}_{ij})$, where $\pi_{ij}=\pi_i + \pi_j$ and $\vec{\mu}_{ij} = {\pi_i \over \pi_{ij}}\vec{\mu}_i + {\pi_j \over \pi_{ij}}\vec{\mu}_j$ (\cite{runnalls2007kullback}). As a result, one can reduce the number of Gaussian components by repeating the above process until it reaches the desired number.  

\section{Experiment setup and result}

\subsection{Stochastic human behavior model training}

\begin{figure} [h]
 \centering
	  \includegraphics[scale=0.6]{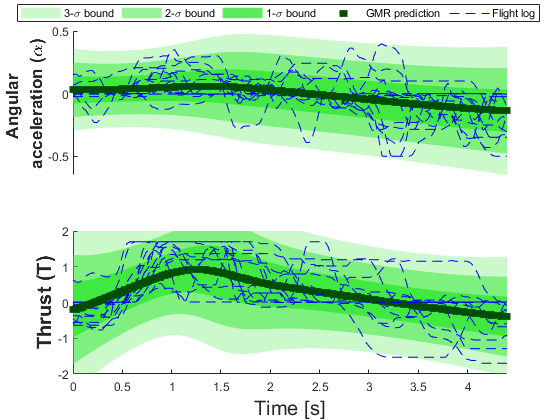}
      \caption{GMR prediction of human control input}
      \label{gmr_fig}
\end{figure}

We first train the stochastic human behavior model using the data obtained from the human subject experiments. A total of 11 participants are recruited from Purdue University\footnote{The Institutional Review Board (IRB) at Purdue University approved the study. IRB protocol number: IRB-2020-755.}. Among the collected data, initial $4.5 \ [s]$ of the trajectory after the obstacle spawned is extracted from each successful landing trial. It yields total $544.5 \ [s]$ of training data from 121 trials. We separate a single trial ($4.5 \ [s]$) as validation data and set $M=3$ for the EM algorithm.

\begin{figure}[tbhp]
\begin{center}
\begin{tabular}[c]{cc}
\begin{subfigure}{.37\textwidth}
\includegraphics[width=\linewidth]{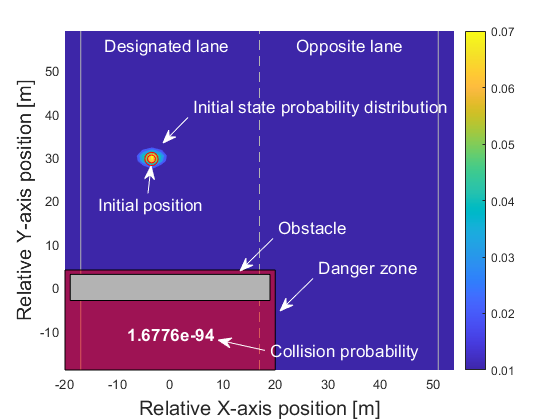}
 \caption{$T=0 \ [s]$, with initial uncertainty}
         \label{fill1}
\end{subfigure} \\

\begin{subfigure}{.37\textwidth}
\includegraphics[width=\linewidth]{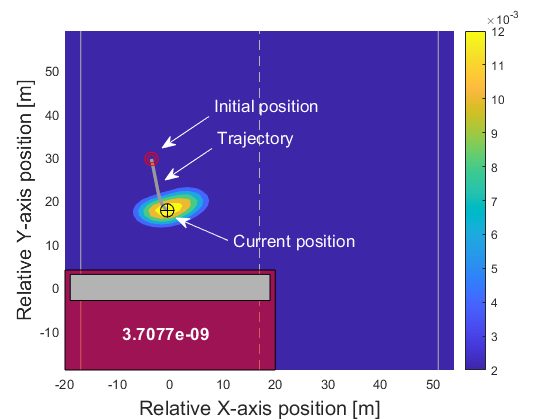}
 \caption{$T=1.5 \ [s]$}
         \label{fill2}
\end{subfigure} \\
\begin{subfigure}{.37\textwidth}
\includegraphics[width=\linewidth]{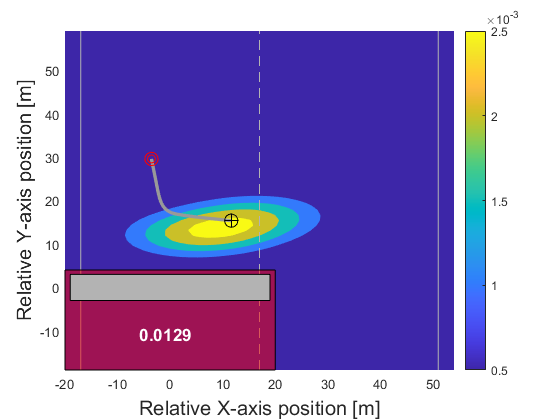}
 \caption{$T=3 \ [s]$}
         \label{fill3}
\end{subfigure} \\

\begin{subfigure}{.37\textwidth}
\includegraphics[width=\linewidth]{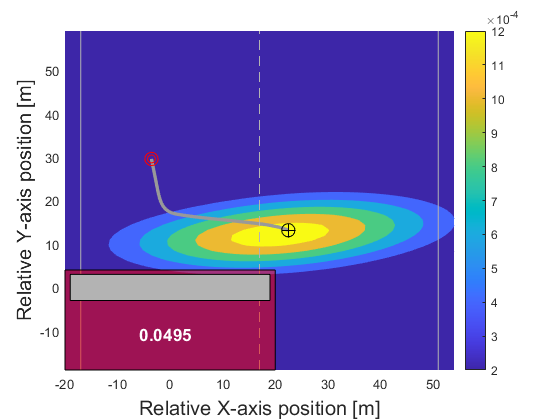}
 \caption{$T=4 \ [s]$}
         \label{fill4}
\end{subfigure}

\end{tabular}
\caption{Comparison between the predicted probability distribution and the true trajectory}
\label{fill_fig}
\end{center}
\end{figure}

\begin{figure}[tbhp]
\begin{center}
\begin{tabular}[c]{cc}
\begin{subfigure}{.37\textwidth}
\includegraphics[width=\linewidth]{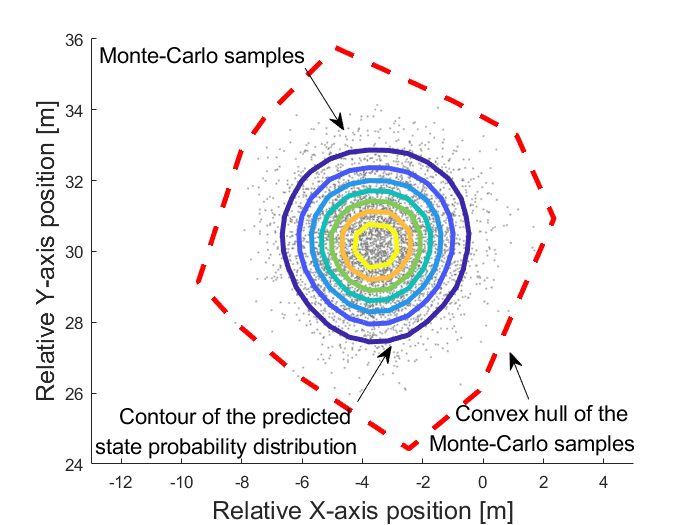}
 \caption{$T=0 \ [s]$}
         \label{cont1}
\end{subfigure} \\

\begin{subfigure}{.37\textwidth}
\includegraphics[width=\linewidth]{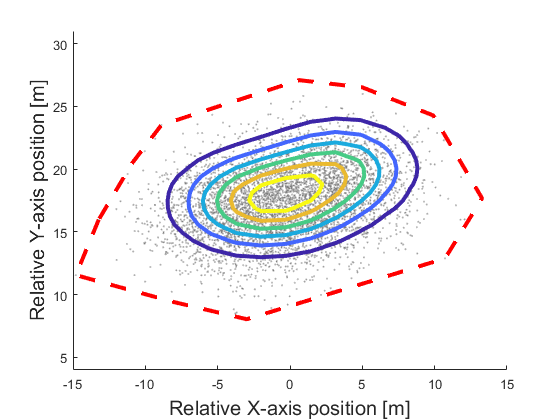}
 \caption{$T=1.5 \ [s]$}
         \label{cont2}
\end{subfigure} \\
\begin{subfigure}{.37\textwidth}
\includegraphics[width=\linewidth]{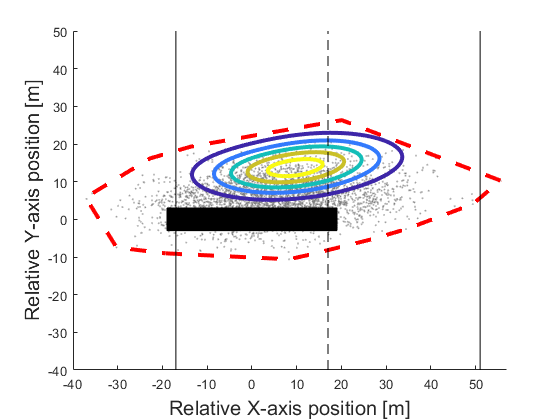}
 \caption{$T=3 \ [s]$}
         \label{cont3}
\end{subfigure} \\

\begin{subfigure}{.37\textwidth}
\includegraphics[width=\linewidth]{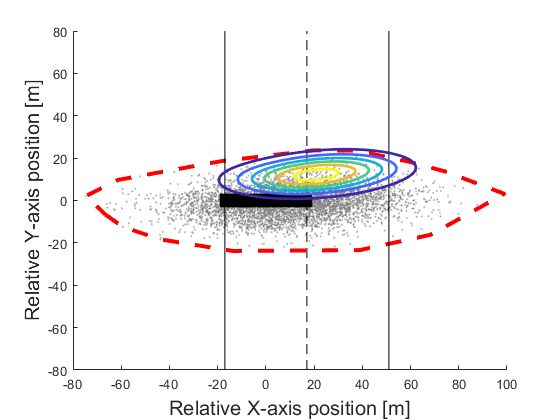}
 \caption{$T=4 \ [s]$}
         \label{cont4}
\end{subfigure}

\end{tabular}
\caption{Comparison between the Monte-Carlo simulation and the proposed state prediction algorithm}
\label{contour_fig}
\end{center}
\end{figure} 

Figure \ref{gmr_fig} shows the randomly selected 10 flight logs and the GMR result for the initial $4.5 \ [s]$ after the obstacle spawned. The green line describes the mean of the GMR result while blue lines are the participants' control input logs. The green shaded regions represent the $1, 2,$ and $3\mhyphen \sigma$, respectively, where $\sigma$ is the standard deviation of the corresponding input. As shown in the figure, the trained model successfully predicts the participants' input within $2\mhyphen \sigma$ bound most of the time.

\subsection{State prediction result}

In this subsection, the state prediction results of the proposed algorithm are presented. Starting from $t_0$, 
the proposed algorithm predicts the states of the multi-rotor at 1.5, 3, and 4 $[s]$. Moreover, the comparison with Monte-Carlo simulation of 5000 samples is also provided to demonstrate the effectiveness of the proposed algorithm.
Table \ref{var} shows the parameters used for the simulation. 
In the simulation, we assume that the initial state has uncertainty represented as a GMM with three Gaussian components.
Accordingly, the initial state, $\vec{x}_0 = [p_{x,0}, p_{y,0}, \theta_{att,0}, v_{x,0}, v_{y,0}, w_0]^T$, with the GMM uncertainty can be written as
\begin{equation} \label{init_gmm}
\bar{\vec{x}}_0 = \vec{x}_0 + \sum^M_{i=1} \pi_{c,i} N(\vec{\mu}_{c,i},\vec{\Sigma}_{c,i}),
\end{equation}
where $\pi_{c,1} = \pi_{c,2} = \pi_{c,3} = {1 \over 3}$, $\vec{\Sigma}_{c,1}$, $\vec{\Sigma}_{c,2}$, and $\vec{\Sigma}_{c,3}$ are identical matrices whose diagonal components are $diag[1.5, 1.5, 0.05, 1, 2, 0.05]$ and $0$ elsewhere, and
\begin{equation} \label{init_unc}
\begin{aligned}
\vec{\mu}_{c,1} &= [0, 0, 0, 0, 0, 0]^T,\\
\vec{\mu}_{c,2} &= [1, 1, 0.1, 1, 1, 0.05]^T,\\
\vec{\mu}_{c,3} &= [-1, 1, -0.1, -1, -1, -0.05]^T.\\
\end{aligned} 
\end{equation} 


\begin{table}[t]
\centering
\captionsetup{width=.6\textwidth}
\caption{Simulation parameters} \label{var}
\begin{tabular}{ | c || c | }
  \hline			
  Discretization time interval ($\Delta t$)  & $0.04 \ [s]$ \\ \hline
  Gravitational acceleration ($g$) & $9.8 \ [m/s^2]$ \\ \hline
  Mass ($m$) & $0.25 \ [kg]$ \\ \hline
  Moment of inertia ($I_x$) & $0.01 \ [kg \cdot m^2]$ \\ \hline
  Control parameter (${k_1, k_2, k_3}$) & ${-0.1, -1, -30}$ \\ \hline
  Thrust input bound ($T$) & $[-1.7, 1.7]$ \\ \hline
  Angular acceleration input bound ($\alpha$) & $[-0.5, 0.5]$ \\ \hline
  Maximum Gaussian components & $16$ \\
  \hline  
\end{tabular} 
\end{table} 

The initial state of each Monte-Carlo simulation is sampled from \eqref{init_gmm} and propagates through the dynamics \eqref{dynamics}. We assume the control inputs ($\alpha, T$) are uniformly distributed with the bound $\alpha \in [-0.5,0.5]$ and $T \in [-1.7,1.7]$ for the Monte-Carlo simulation, as shown in Table \ref{var}.

Figure \ref{fill_fig} shows the state prediction result at each specific time instant. In the figures, the grey box is the obstacle, the red circle is the initial position, the grey line is the actual trajectory of the multi-rotor, and the black circle is the position of the multi-rotor at the given time instant. The white lines at each side of the figures and the dotted line at the center divide the lanes, which are described in Fig. \ref{scenario_fig}. The red box at the bottom of the figures is the danger zone where the multi-rotor should avoid. The collision probability is obtained by computing the cumulative probability of the predicted probability distribution on the danger zone. Figure \ref{fill1} describes the initial state with the uncertainty given in \eqref{init_unc}. Meanwhile, Figs. \ref{fill2}-\ref{fill4} represent the predicted state probability distribution at each time instant. The brighter area means higher probability. Due to the initial uncertainty and the stochastic human behavior model, the probability distribution expands over time. 
Nevertheless, the additional probabilistic information gives more weight on a certain area unlike the conventional reachable set.

Figures \ref{cont1}-\ref{cont4} are the comparison between the proposed algorithm and the Monte-Carlo simulation. The contour describes the predicted probability distribution, the grey dots are the Monte-Carlo samples, and the red dotted line is the convex hull of the samples. The black lines and box represent the lanes and the obstacle, respectively. In the figures, the resulting contour significantly reduces the area that the multi-rotor is likely to reach in comparison with the convex hull. This advantage is clearly shown in $3$-$4 \ [s]$. In Figs. \ref{cont3}-\ref{cont4}, the Monte-Carlo samples indicate that the multi-rotor can reach the obstacle although there is no collision. In contrast, the proposed approach shows in Figs. \ref{fill3}-\ref{fill4} that the multi-rotor is not likely to collide, thereby alleviating the conservativeness of reachable states, a well-known weakness of the reachability analysis. 

\section{Conclusion}

In this paper, we proposed the state prediction algorithm for the Human-in-the-Loop Cyber Physical System (HiLCPS) which explicitly considers the stochastic human behavior model. The proposed algorithm can compute the state probability distribution of the multi-rotor at a given future time instant. The computed probability distribution provides more information that can be facilitated for safe controller design compared to a simple set of states which further improves the reliability of the system. 

For future work, the application of the proposed algorithm to various systems will be considered, such as the safety envelope analysis of a car. The flexibility of the proposed algorithm will be demonstrated by transferring the algorithm to other systems.


\bibliography{ifacconf}             
                                                   







\end{document}